\pdfoutput=1

\documentclass[%
 reprint,
 superscriptaddress,
 nobibnotes,
 amsmath,amssymb,
 aps,
 showkeys,
 longbibliography,
]{revtex4-2}

\usepackage{graphicx}
\usepackage{dcolumn}
\usepackage{bm}
\usepackage{chemformula}
\usepackage[separate-uncertainty=true,separate-uncertainty,multi-part-units=single]{siunitx}
\usepackage{hyperref}
\hypersetup{colorlinks=true,allcolors=magenta}

\begin{document}

\title{Understanding the Role of Open Metal Sites in MOFs for the Efficient Separation of Benzene/Cyclohexane Mixtures}

\author{C. González-Galán}
    \affiliation{Department of Physical, Chemical, and Natural Systems, Universidad Pablo de Olavide. Ctra. Utrera km 1. ES-41013 Seville, Spain}
\author{R. M. Madero-Castro}
    \affiliation{Department of Physical, Chemical, and Natural Systems, Universidad Pablo de Olavide. Ctra. Utrera km 1. ES-41013 Seville, Spain}
\author{A. Luna-Triguero}
    \affiliation{Energy Technology, Department of Mechanical Engineering, Eindhoven University of Technology, P. O. Box 513, 5600 MB Eindhoven, The Netherlands}
    \affiliation{Eindhoven Institute for Renewable Energy Systems (EIRES), Eindhoven University of Technology, Eindhoven 5600 MB, The Netherlands}
\author{J. M. Vicent-Luna}
    \email[Corresponding author: ]{j.vicent.luna@tue.nl}
    \affiliation{Materials Simulation \& Modelling, Department of Applied Physics, Eindhoven University of Technology, 5600 MB, Eindhoven, The Netherlands}
    \affiliation{Eindhoven Institute for Renewable Energy Systems (EIRES), Eindhoven University of Technology, Eindhoven 5600 MB, The Netherlands}
\author{S. Calero}
    \email[Corresponding author: ]{s.calero@tue.nl}
    \affiliation{Materials Simulation \& Modelling, Department of Applied Physics, Eindhoven University of Technology, 5600 MB, Eindhoven, The Netherlands}
    \affiliation{Eindhoven Institute for Renewable Energy Systems (EIRES), Eindhoven University of Technology, Eindhoven 5600 MB, The Netherlands}

\date{\today}

\begin{abstract}

Separating C6 cyclic hydrocarbons, specifically benzene and cyclohexane, presents a significant industrial challenge due to their similar physicochemical properties. We conducted Monte Carlo simulations in the Grand-Canonical ensemble to acquire adsorption properties and separation performance data for benzene and cyclohexane in three metal-organic frameworks featuring coordinatively unsaturated metal sites (Ni-MOF-74, Ni-ClBBTA, and Ni-ClBTDD). The separation performance of these MOFs was analyzed and compared with literature data for adsorbents of different natures, demonstrating superior performance. Additionally, we explored the molecular origins of this effective separation, examining the pore-filling mechanism, interaction of guest molecules with metal centers, and mutual interactions of each adsorbate. Our results highlight that the selected adsorbents, with remarkable loading capacity, can efficiently separate both compounds in a mixture with exceptional effectiveness.

\end{abstract}


\maketitle


\section{Introduction}
\label{sec:intro}
Benzene is an aromatic hydrocarbon we find naturally in the environment at low concentration. It is widely used in several processes in the chemical industry, pointing out its use in the production of polymers, resins, and synthetic fibers.\cite{Galbraith2010} However, many reports about the toxicity of benzene can be found in the literature. It is well known that benzene is a genotoxic carcinogen.\cite{Shallis2021, Dewi2019, Loomis2017} Inhalation of benzene can cause different effects depending on the concentration and the time of exposure. Leukemia, aplastic anemia, vertigo, drowsiness, headache, or nausea are some of the toxic effects of this organic compound.\cite{DuarteDavidson2001} For this reason, capture and separation of benzene are required. Separation and purification are one of the most important processes in the chemical industry. Of critical importance is the separation of C6 cyclic hydrocarbons benzene and cyclohexane. C6 cyclic hydrocarbons separation is a complex process due to the similarities between benzene and cyclohexane. These organic compounds have very close boiling points, similar molecular volumes, and geometries, and they also form an azeotropic mixture with a content of 45\% cyclohexane.\cite{GarciaVillaluenga2000, Yildirim2008, Yamasaki1997}

In recent years, the use of porous materials for benzene-cyclohexane separation has increased over conventional methods such as distillation, liquid-liquid extraction, or membrane pervaporation.\cite{Mukherjee2021} Although there are some studies that evidence the similar adsorption of benzene and cyclohexane in many frameworks,\cite{Wang2015, Crespo2006, Eddaoudi2000} adsorption-based separation using porous materials is a great alternative to conventional methods. Metal-organic frameworks (MOFs), porous organic polymers (POPs), zeolites and their derivatives have been used for this purpose. González-Galán et al.\cite{GonzlezGaln2020} studied the targeted separation using aluminosilicates (FAU zeolite modified with Na cations) based on the interaction between the electronic ring of benzene and the cation obtaining a very high separation. In the same way, Min Tu et al.\cite{Tu2015} described the separation of this pair of molecules using zeolitic-imidazolate frameworks (ZIFs) with different metal centers. Based on the interaction between open metal sites and benzene, Mukherjee et al.\cite{Mukherjee2016} studied the M-MOF-74 family (with different metal centers, M) to perform the separation. Similarly, Manna et al.\cite{Manna2015} and Cheng et al.\cite{Cheng2014} used DAT-MOF-1 and Ag(I)- MOF, respectively, to carry out the separation. Some studies using POPs have also been published,\cite{Deng2017, Yan2016, Li2014} but the absence of a specific interaction site reduces the separation performance. Mukherjee et al.\cite{Mukherjee2021} reviewed the progress of the adsorption-based benzene/cyclohexane separation using MOFs, discussing the specific interactions between adsorbents and adsorbates that affect the separation performance. In the same work, the authors unraveled the particular interactions responsible for the inverted selectivity of cyclohexane over benzene in MOF-5 by means of DFT calculations. Recently, Jansen et al. \cite{Jansen2022} studied the performance of 18 MOFs for the targeted separation, aimed to link the structural properties of the adsorbents with the separation performance or the benzene adsorption capacity.

Metal-organic frameworks are a class of crystalline porous materials containing metal centers and organic linkers. Due to their structural and functional tunability, they have a wide variety of applications including separation and purification, catalysis and/or drug delivery, among others.\cite{Kotzabasaki2018, Lin2019, Yang2019} Within the family of MOFs, some of them stand out due to the accessibility of the metallic centers (open metal site, OMS), which may affect some of their applications. For instance, the capture of CO$_2$ using Mg-MOF-74 is favored by the interaction of the molecule with the open metal site.\cite{Yazaydn2009} This interaction is stronger for molecules with double or triple bonds, such as olefins, than for other small gases or paraffins.\cite{LunaTriguero2017, LunaTriguero2017a, LunaTriguero2018, Fischer2012, Jorge2014, Karra2008, Dietzel2009, Yazaydn2009a} This particular interaction between an open metal center and molecules with double or triple bonds is known as $\pi$-complexation and is not only present in MOFs with open metal sites, but it also happens in some aluminosilicates.\cite{GonzlezGaln2020, LunaTriguero2020, Yang1995, Khelifa1999} Liu et al. studied the performance and the molecular mechanisms involved in the separation of benzene and cyclohexane in Mg-MOF-74.\cite{Liu2019} In this work, we exploit the interaction between benzene and metal centers in MOFs with OMS for the separation of benzene and cyclohexane.

Ni-MOF-74 is one of the components of the M-MOF-74 family. It is formed by the coordination of metallic centers (Ni) and organic ligands (2,5-dihydroxybenzenedicarboxilyc acid). It stands out due to the presence of open metal sites in its 1-dimensional hexagonal pore channel. In the same way, Ni-ClBBTA and Ni-ClBTDD show the same topology but with different pore dimensions (Ni-ClBTDD exhibits the biggest pore size, approximately 18 Å). The presence of open metal sites in their structures and the extremely high loading capacity make them promising materials for the targeted separation. Here, we study three MOFs - Ni-MOF-74, Ni-ClBTDD, and Ni-ClBBTA - for the separation of benzene and cyclohexane based on the interaction between the aromatic ring of benzene and the metallic centers. The selected MOFs have the advantage that they are very stable structures in terms of humidity and temperature. Moreover, their loading capacity is bigger than other studied porous materials. For this purpose, we have used Monte Carlo simulations to analyze the separation performance of the three selected adsorbents.

\section{Methodology}
\label{sec:methods}
Adsorbates were described using united atom rigid models, considering the CH and CH$_2$ groups as single pseudo-atoms. In the case of cyclohexane, we used two conformers – chair and t-boat – which are the most stable configurations of the molecule.\cite{Sawek2018} Note that the total adsorption loading is the sum of both. In the case of benzene, we used a nine-site model of the TraPPE-UA force field.\cite{Rai2012} In this model, the molecule of benzene is described considering that each CH group is a single Lennard-Jones center, and three partial charges are assigned. The first one is a positive charge placed at the center of the ring and the other charges are two negative ones representing the electronic cloud of benzene and they are placed perpendicularly to the plane of the aromatic ring. 
The Lennard-Jones parameters were taken from Universal Force Field (UFF) for metallic centers and DREIDING force field for organic linkers.\cite{Rappe1992, Mayo1990} Host-guest, and guest-guest interactions were described using Lorentz-Berthelot mixing rules. This procedure can be found in many reports. Metal-organic frameworks used in the present work and their corresponding pore size distribution are shown in Figure S1. All of them point out by the presence of open metal site and differ in the size of their pores. Charges can be found in the supporting information (Figure S2).
Configurational-Bias Monte Carlo (CBMC) in the Grand-Canonical ensemble (µVT) was used to compute single and multicomponent adsorption isotherms at 298K. In this ensemble, the fluctuation of the molecules to reach equilibrium is permitted by fixing the chemical potential (µ), the volume (V), and the temperature (T).\cite{Dubbeldam2013} It is worth noting that for cyclohexane, we maintain equimolar amounts of chair and t-boat conformers in the reservoir, and the adsorption depends on the affinity. All the simulations performed in this work have been carried out using the RASPA code.\cite{Dubbeldam2015} 

The average occupation profiles were computed by averaging the entire trajectory recorded during Grand-Canonical Monte Carlo (GCMC) simulations at a selected value of pressure. This allows us to know molecular configurations of the adsorbates inside the frameworks. The $\pi$-interaction between the pure adsorbates (molecules of benzene) was computed by taking into account all the possible aromatic stacking arrangements and the geometrical criteria associated to each one.\cite{Martinez2012, Dance2002} 

To assess the separation performance of the adsorbents for the targeted application, we used the following metrics. Firstly, we process the adsorption equilibrium data generated by the GCMC simulations and data from experiments reported in the literature by fitting them to an adsorption equation. We used the Dual-site Langmuir Sips \cite{sips1948} equation:

\begin{equation} \label {eq1}
			q(P)=\sum_i q_i^{sat}\frac{(bp)^{1/\nu_i}}{1+(bp)^{1/\nu_i}}
	\end{equation}

The competitive adsorption of a gas mixture is characterized by the working capacity of each component at a given condition. The working capacity represents the loading difference between the adsorption and desorption operating conditions:

\begin{equation} \label {eq2}
        \Delta q= q_{ads}-q_{des}= q(P_{ads},T_{ads})-q(P_{des},T_{des})
	\end{equation}

\noindent where $P_{ads},T_{ads},P_{des},$ and $T_{des}$ are the adsorption and desorption condition, respectively.

\begin{figure*}[!t]
\begin{center}
\includegraphics[width=0.75\linewidth]{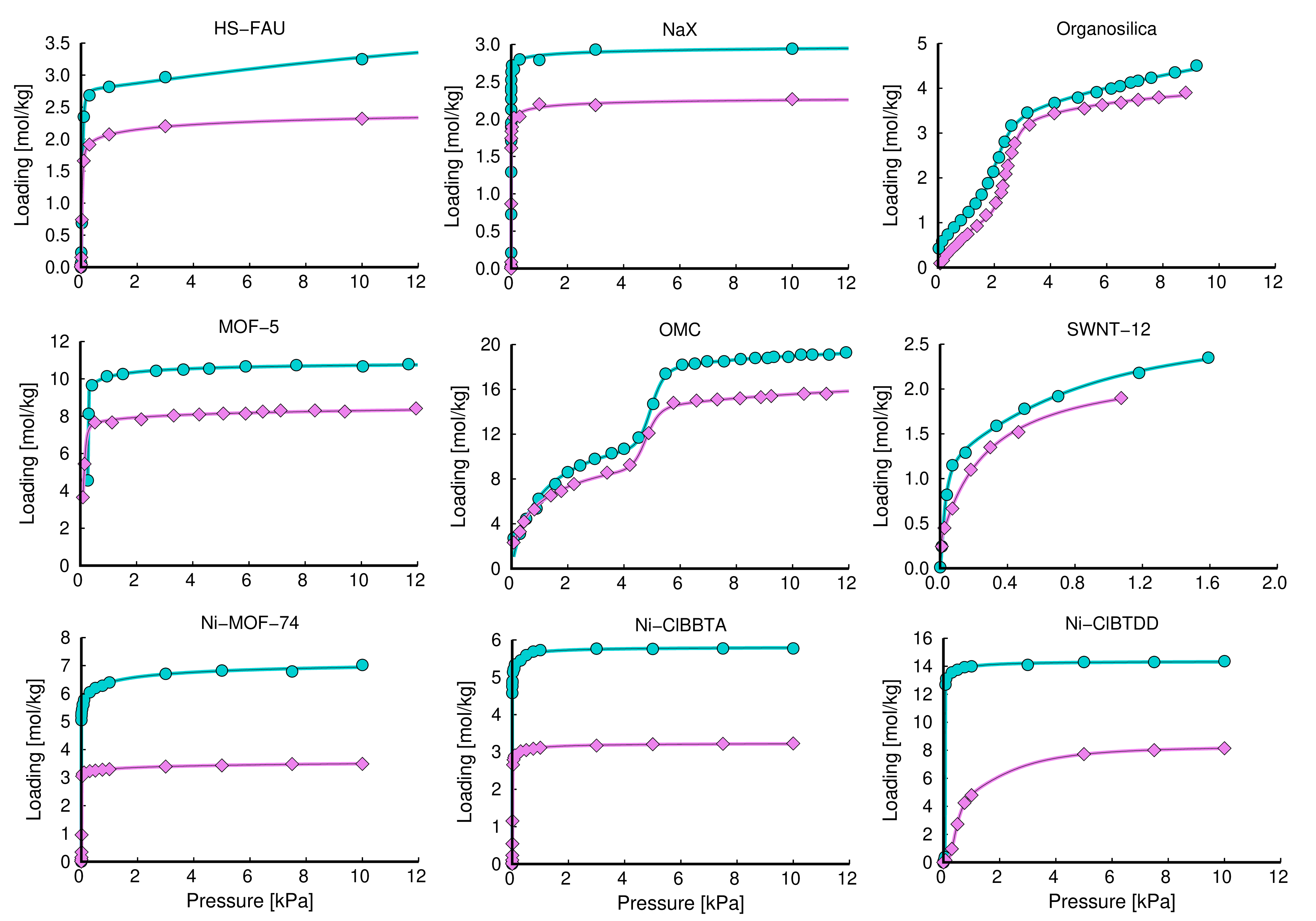}
\caption{Single adsorption isotherms of benzene (cyan) and cyclohexane (pink) in several porous materials: HS-FAU and NaX zeolites,\cite{GonzlezGaln2020} mesoporous organosilica,\cite{Wu2021} MOF-5,\cite{Eddaoudi2000a} ordered mesoporous carbons,\cite{Wang2015a} carbon nanotubes 12 Å,\cite{Crespo2006a} Ni-MOF-74, Ni-ClBBTA, and Ni-ClBTDD. The lines correspond to the fitting to the dual-site Langmuir Sips isotherm. Table S1 collects the parameters of the fitting.}
\label{label:fig_01}
\end{center}
\end{figure*}

Then, we defined the productivity of a mixture separation as the ratio of the working capacities of each component divided by the total working capacity:

\begin{equation} \label {eq3}
        P_i= \frac{\Delta q_i}{\Delta q_{all}}
	\end{equation}

 \noindent where $\Delta q_{all}=\sum_i \Delta q_i$

Assuming complete regeneration of the material, i.e., desorption conditions leading to the complete desorption of all components of the mixture, the productivity is reduced to the molar fraction of each component at the adsorption conditions. This scenario represents a vacuum swing adsorption process. 

Next, we computed the competitive adsorption of benzene and cyclohexane using IAST \cite{IAST-1965} module implemented in the RUPTURA code.\cite{RUPTURA} The adsorption selectivity is defined as:

\begin{equation} \label {eq4}
        S_{ads}(i/j)= \frac{y_i/y_j}{x_i/x_j}
	\end{equation}

\noindent where $y$ is the molar fraction in the adsorbed phase, and $x$ is the molar fraction in the bulk phase of components $i$ and $j$, respectively. For the case of an equimolar mixture, the adsorption selectivity is reduced to the ratio of the loading of benzene and cyclohexane in the mixture. Because of the definition of the adsorption selectivity, situations where one of the adsorbates is almost excluded from a mixture lead to remarkably high selectivity values. Hence, the comparison of the performance of different adsorbents becomes difficult. Therefore, we used a separation factor based on a trade-off between selectivity and adsorption capacity to reduce the impact of those high selectivity values.

\begin{equation} \label {eq5}
        \alpha_(i/j)=ln(S_{ads}(i/j)) \cdot \Delta q_i
	\end{equation}

This metric has been reported in the literature \cite{Zhiwei2018,Shah2016,Gao2023} and proves to be an adequate indicator of the separation performance of a gas mixture.

\section{Results and Discussion}
\label{sec:results}

\begin{figure}[!t]
\begin{center}
\includegraphics[width=1.0\linewidth]{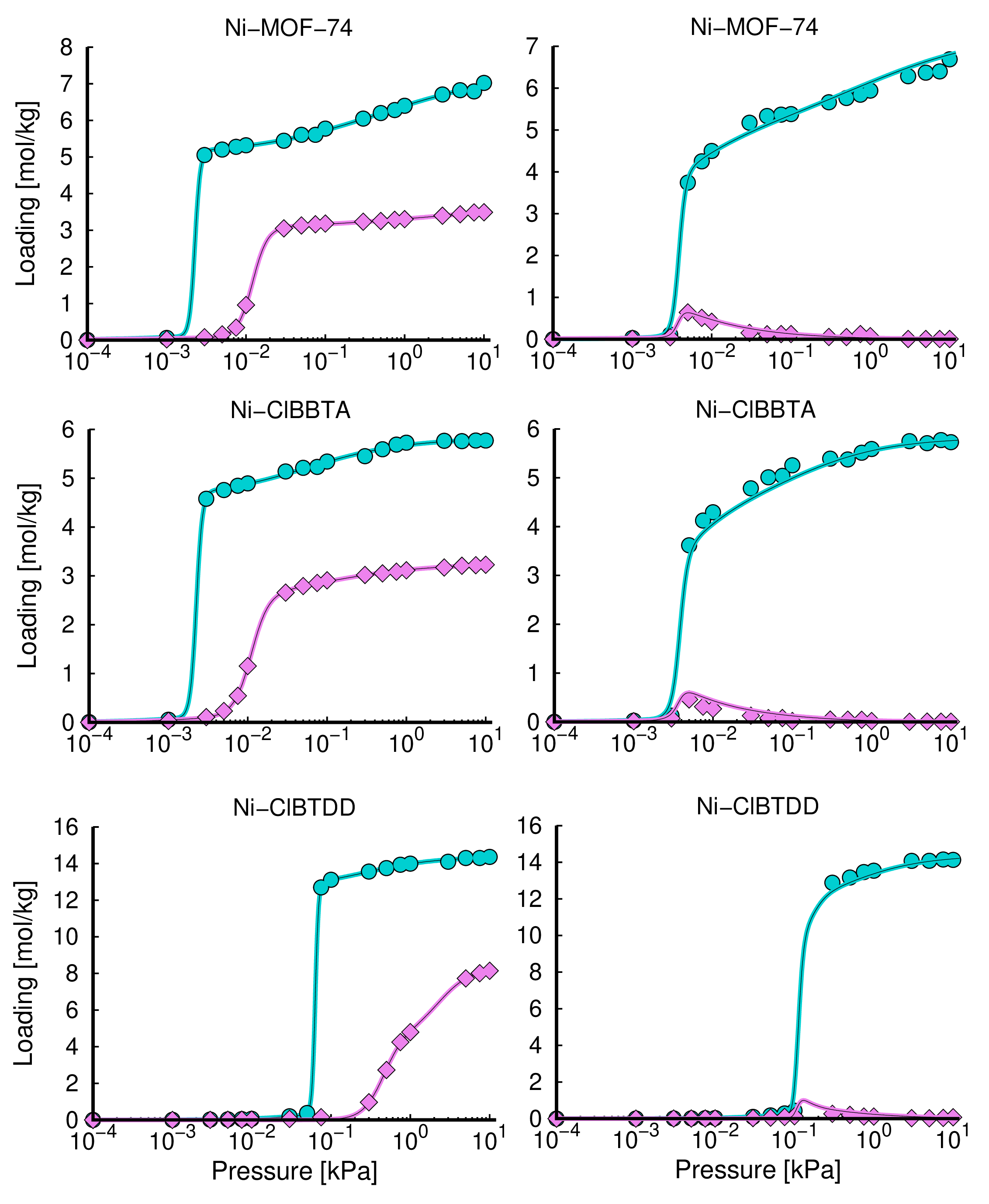}
\caption{Single (left) and multicomponent adsorption (right) isotherms of benzene (cyan) and cyclohexane (pink) in Ni-MOF-74, Ni-ClBBTA, and Ni-ClBTDD at 298 K. For pure component adsorption isotherms, the solid lines represent the fitting to the to the dual-site Langmuir Sips isotherm, while for multicomponent adsorption, the solid lines represent the IAST predictions of the binary mixture.}
\label{label:fig_02}
\end{center}
\end{figure}

Despite the high industrial interest in the separation of benzene and cyclohexane, this separation has not been extensively studied in the literature. Figure \ref{label:fig_01} presents the single adsorption isotherms of these two molecules in various porous materials of different natures, such as zeolites, carbon nanotubes, mesoporous carbons, organosilica materials, and metal-organic frameworks.\cite{GonzlezGaln2020, Wang2015a, Crespo2006a, Eddaoudi2000a, Wu2021} The selected adsorbent set covers a wide range of materials, all with a pore size larger than 10 Å, aiming to maximize the uptake of these compounds. The set also includes two mesoporous materials, an ordered mesoporous carbon and a mesoporous organosilica.

Among the adsorbents depicted in Figure \ref{label:fig_01}, we observe that MOFs can discriminate between these similar molecules, also exhibiting high adsorption capacity. In particular, Ni-MOF-74, Ni-ClBBTA, and Ni-ClBTDD, which contain open metal sites, serve as preferential adsorption sites for molecules with double bonds.\cite{LunaTriguero2017, LunaTriguero2017a, LunaTriguero2018, Fischer2012, Jorge2014, Karra2008, Dietzel2009, Yazaydn2009a, Liu2019} MOF-5 shows a higher adsorption capacity of benzene in the high-pressure range, while cyclohexane exhibits higher affinity in the lower pressure range, leading to a reverse selectivity, as analyzed by Mukherjee et al.\cite{Mukherjee2021} Aluminosilicates (HS-FAU and NaX) are also suitable adsorbents for this separation due to the strong electrostatic interaction between benzene and the extraframework cations.\cite{GonzlezGaln2020, LunaTriguero2020} However, zeolites exhibit a lower adsorption capacity than MOFs. Finally, the adsorption behavior of benzene and cyclohexane is quite similar in carbon nanotubes, mesoporous carbons, or organosilica materials.

To assess the capacity of the chosen materials for separating benzene from cyclohexane, we conducted a study on multicomponent competitive adsorption using the Ideal Adsorbed Solution Theory (IAST). Initially, to validate the predictive accuracy of IAST for competitive adsorption of these two molecules, we computed the adsorption isotherms for pure components and an equimolar mixture at room temperature for Ni-MOF-74, Ni-ClBBTA, and Ni-ClBTDD (Figure \ref{label:fig_02}). The noticeable difference in onset pressures of adsorption for the individual components indicates a higher affinity of this MOF family for benzene compared to cyclohexane. Furthermore, the amount of adsorbed benzene in the three MOFs is nearly double that of cyclohexane. These factors significantly influence the separation performance of the mixture, as detailed below. The analysis of the multicomponent adsorption isotherms (Figure \ref{label:fig_02}) reveals a consistent trend of increased benzene adsorption compared to cyclohexane, excluding the latter in the case of the equimolar mixture. Another noteworthy observation is the higher storage capacity of Ni-ClBTDD compared to the other MOFs, providing an additional advantage over other porous adsorbent materials. The strong agreement between IAST predictions and GCMC simulations for the benzene/cyclohexane equimolar mixture not only affirms the validity of IAST for studying this specific mixture but also extends its applicability to other adsorbents reported in the literature.  

\begin{figure}[!t]
\begin{center}
\includegraphics[width=0.8\linewidth]{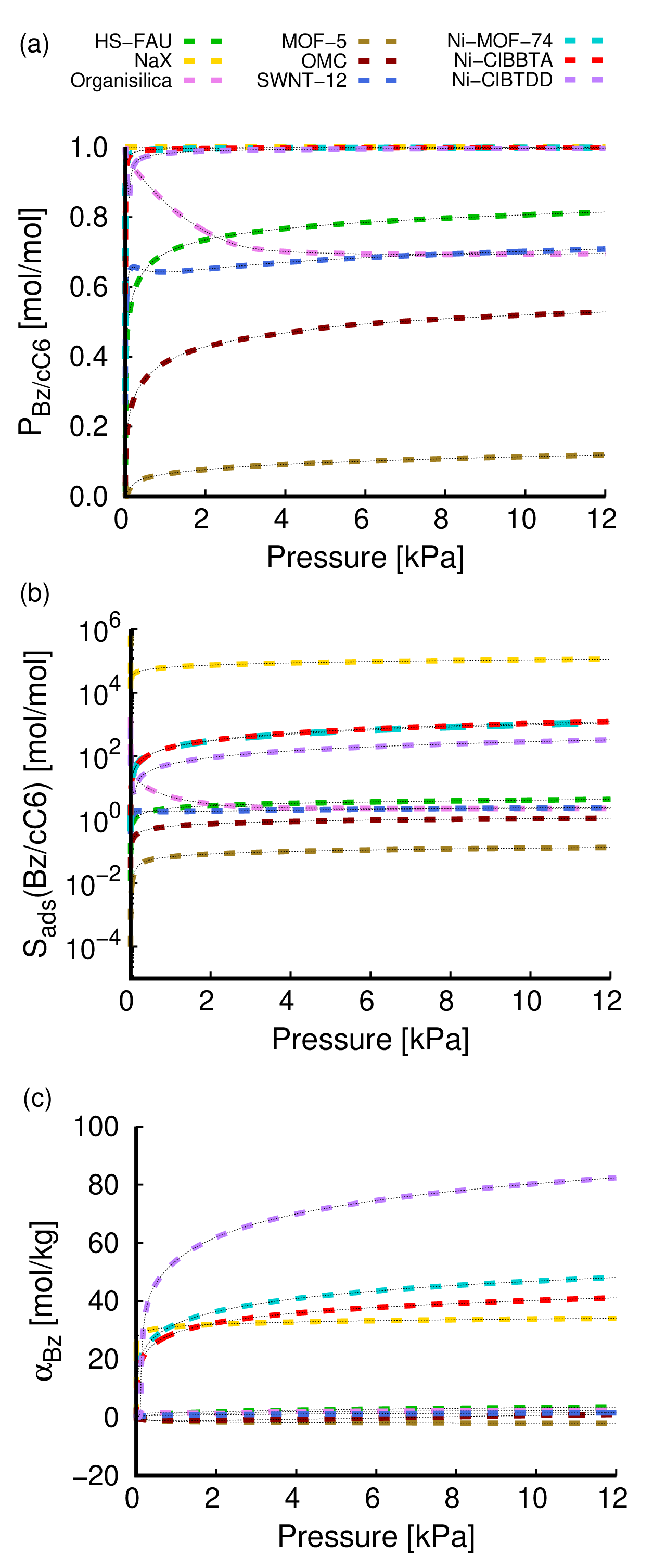}
\caption{P$_{(Bz/cC6)}$, Productivity of the separation (a), S$_{ads(Bz/cC6)}$, adsorption selectivity (b), and $\alpha_{Bz}$, separation factor as the trade-off between selectivity and adsorption capacity (c) for the equimolar mixture of benzene and cyclohexane in the selected adsorbents at 298 K.}
\label{label:fig_03}
\end{center}
\end{figure}

\begin{figure}[!t]
\begin{center}
\includegraphics[width=0.8\linewidth]{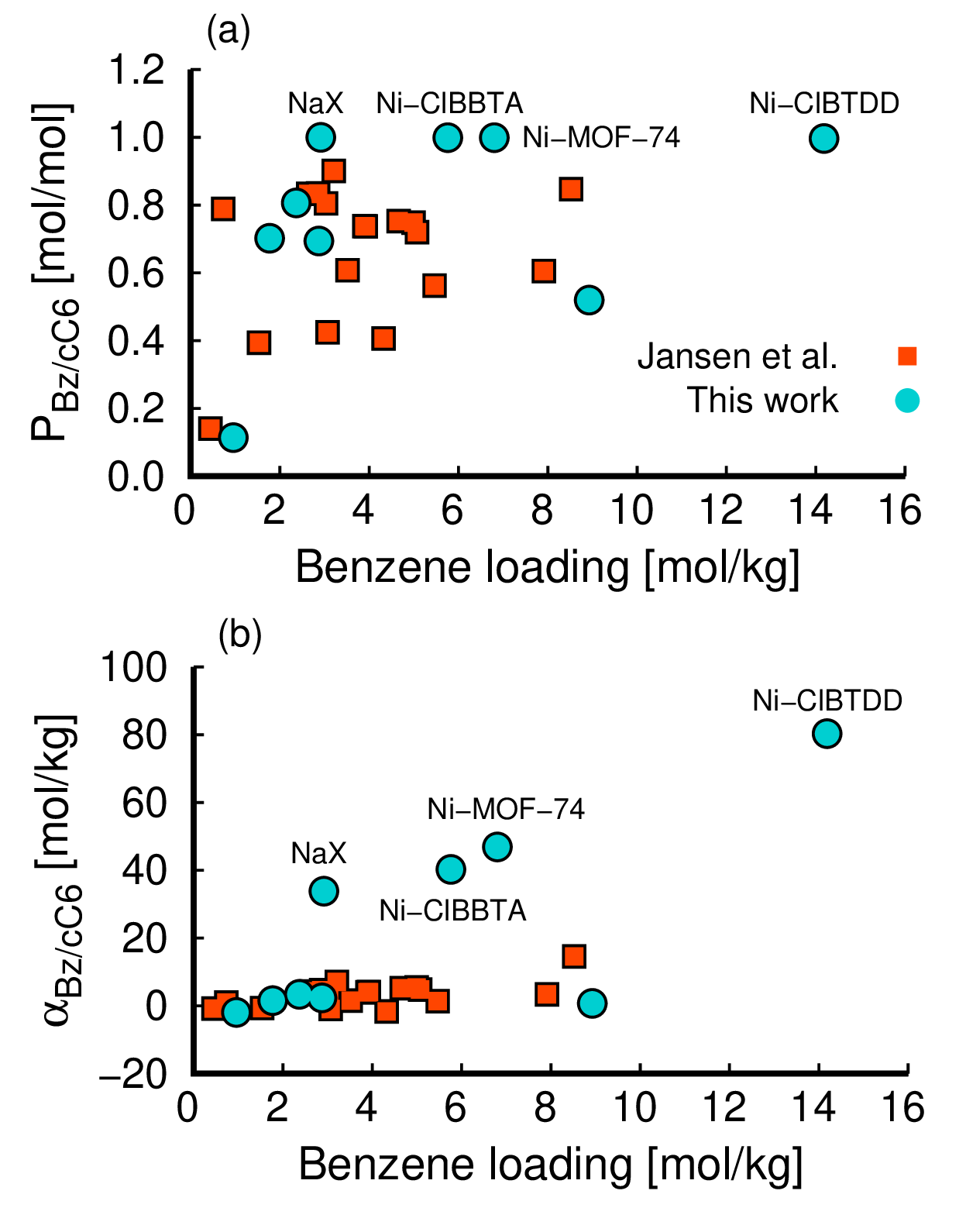}
\caption{P$_{(Bz/cC6)}$, Productivity of the separation (a), S$_{ads(Bz/cC6)}$, and $\alpha_{Bz}$, separation factor as the trade-off between selectivity and adsorption capacity (b) for the equimolar mixture of benzene and cyclohexane in the selected adsorbents at 10 kPa and 298 K. The figure also includes the values calculated from the adsorption equilibrium reported by Jansen et al.\cite{Jansen2022} at 293 K. The values corresponding to each symbol in the figures are collected in Tables S2 and S3, respectively.}
\label{label:fig_04}
\end{center}
\end{figure} 

\begin{figure}[!t]
\begin{center}
\includegraphics[width=0.98\linewidth]{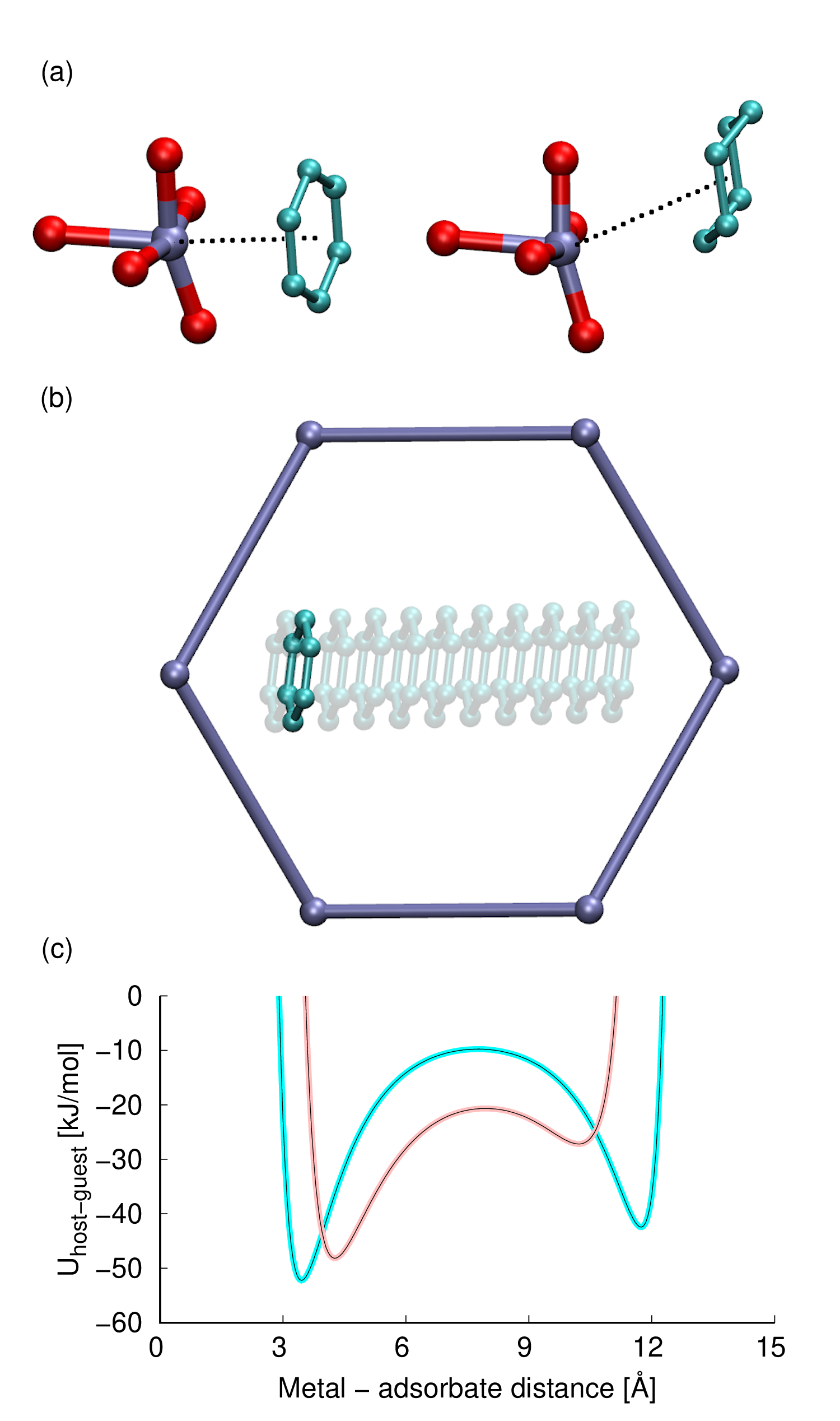}
\caption{Binding geometries of benzene and cyclohexane adsorbed near the OMs (a), and schematics (b) and host-guest potential energy of displacements (c) of guest molecules starting from their equilibrium configurations. The displacements of the adsorbates are linear translations of the center of mass of the molecules with respect to the metal center.}
\label{label:fig_05}
\end{center}
\end{figure} 

To assess the separation performance of benzene and cyclohexane in multicomponent adsorption, detailed metrics were analyzed as outlined in the Methodology section. These metrics included the productivity of separation, adsorption selectivity, and the separation factor, representing the trade-off between selectivity and adsorption capacity. Figure \ref{label:fig_03} illustrates these metrics for the equimolar mixture of benzene and cyclohexane across all the adsorbents depicted in Figure \ref{label:fig_01}. Figure \ref{label:fig_03}a displays the productivity of the mixture under the assumption of full adsorbent regeneration, where operating conditions lead to the total desorption of both compounds. Four adsorbents, namely NaX zeolite, and the three MOFs with open metal sites (Ni-MOF-74, Ni-ClBBTA, and Ni-ClBTDD), exhibit nearly complete exclusion of cyclohexane from the mixture, with a productivity factor close to unity. A previous study highlighted the exceptional separation performance of NaX,\cite{GonzlezGaln2020} attributed to the interaction between benzene and the extraframework sodium cations in this aluminosilicate. Similar interactions between the open metal sites of large cavity MOFs and the aromatic ring of benzene contribute to high separation from cyclohexane.\cite{Li2014, Mukherjee2016, Mukherjee2021}

While adsorption selectivity is a common measure for evaluating mixture separation, its reliance on the ratio of component loadings can lead to extremely high values when one component is nearly excluded. Figure \ref{label:fig_03}b demonstrates this issue, with the four materials excluding cyclohexane showing exceptionally high adsorption selectivity values, with NaX being the most promising adsorbent. However, these extremely high values may be influenced by the division of very small numbers, making selectivity less ideal factor for assessing separation performance. To address this limitation, we recommend using a separation factor, representing the trade-off between adsorption selectivity and capacity (eq. \ref{eq5}). Figure \ref{label:fig_03}c represents this separation factor for all studied adsorbents. Remarkably, the same four outstanding materials from Figures 4a and 4b emerge as the top candidates. However, in this context, Ni-ClBTDD demonstrates a substantial improvement over Ni-MOF-74, Ni-ClBBTA, and NaX zeolite. This is attributed to Ni-ClBTDD exhibiting not only high separation performance but also a significantly greater benzene adsorption capacity compared to the rest of the adsorbents (see Figure \ref{label:fig_01}).

The evaluation of separation performance metrics in Figure \ref{label:fig_03} readily extends to the assessment of gas mixture separation, starting from the adsorption equilibrium of pure components. To demonstrate the exceptional efficiency of MOFs with open metal sites in separating benzene and cyclohexane, we compared productivity and separation factors with all the materials selected in this study, along with MOFs reported by Jansen et al.\cite{Jansen2022} In Figure \ref{label:fig_04}, productivity and separation factors at 10 kPa are depicted as a function of the loading of benzene.

Figure \ref{label:fig_04} unequivocally illustrates the superior performance of large cavity MOFs with open metal sites over other proposed adsorbents in the literature. Among these, Ni-ClBTDD emerges as the most effective adsorbent for benzene capture and purification. This comparison highlights the exceptional potential of MOFs with open metal sites, emphasizing their applicability for enhancing the separation efficiency of benzene and cyclohexane mixtures in various industrial processes.

\begin{figure}[!t]
\begin{center}
\includegraphics[width=0.98\linewidth]{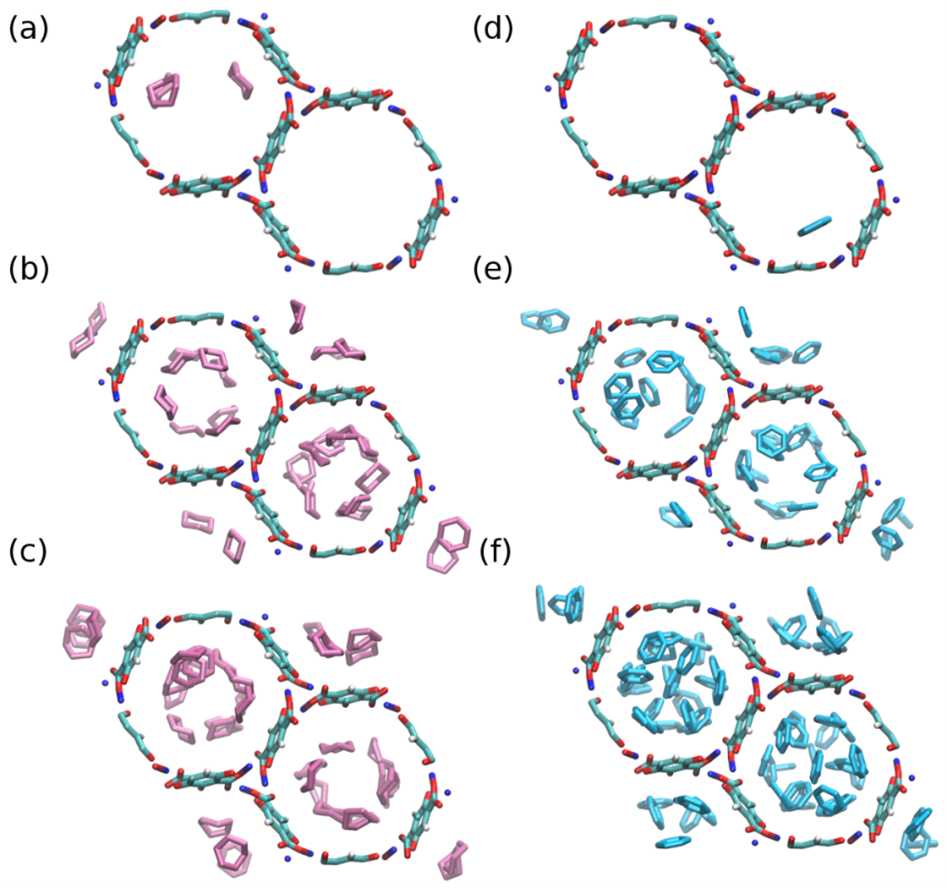}
\caption{Representative snapshot of the occupation of molecules of cyclohexane (left) and benzene (right) inside Ni-MOF-74 at (a, d) low, (b, e) medium, and (c, f) high loading at 298K.}
\label{label:fig_06}
\end{center}
\end{figure}

Understanding the separation process implies the knowledge of the interaction of the adsorbates with the adsorbents and the adsorbates between them. The interaction of isolated molecules with the adsorbents governs the infinite dilution regime. Figure \ref{label:fig_05}a shows the binding geometries of benzene and cyclohexane adsorbed in the metal cluster of Ni-MOF-74. Benzene molecules are located in a parallel conformation to the plane formed by the oxygen atoms attached to the metal center where the center of the aromatic ring faces the Ni metal atoms. Cyclohexane molecules are also located close to the metal centers but with a different alignment than benzene. While the distance between the metal atoms and the center of mass of the molecules is 3.45 Å for benzene, this value is about 4.5 Å for cyclohexane. Figure \ref{label:fig_05} also shows the potential interaction energy of the molecules as displaced from their equilibrium configurations. These energy profiles show that the repulsive energy barrier for cyclohexane takes place at long distances to the metal center compared to benzene and, therefore, to the internal surface of the MOF. The energy profiles show an additional minimum and a secondary repulsive barrier as increasing the interatomic distance when the molecules approach the opposite wall of the cavities. The shorter distances between the repulsive energy barriers suggest that cyclohexane has less accessibility to the available pore, thus reducing the adsorption uptake, even at pressure values close to the vapor pressure of the adsorbates (see Figure \ref{label:fig_02}).

\begin{figure}[!t]
\begin{center}
\includegraphics[width=0.98\linewidth]{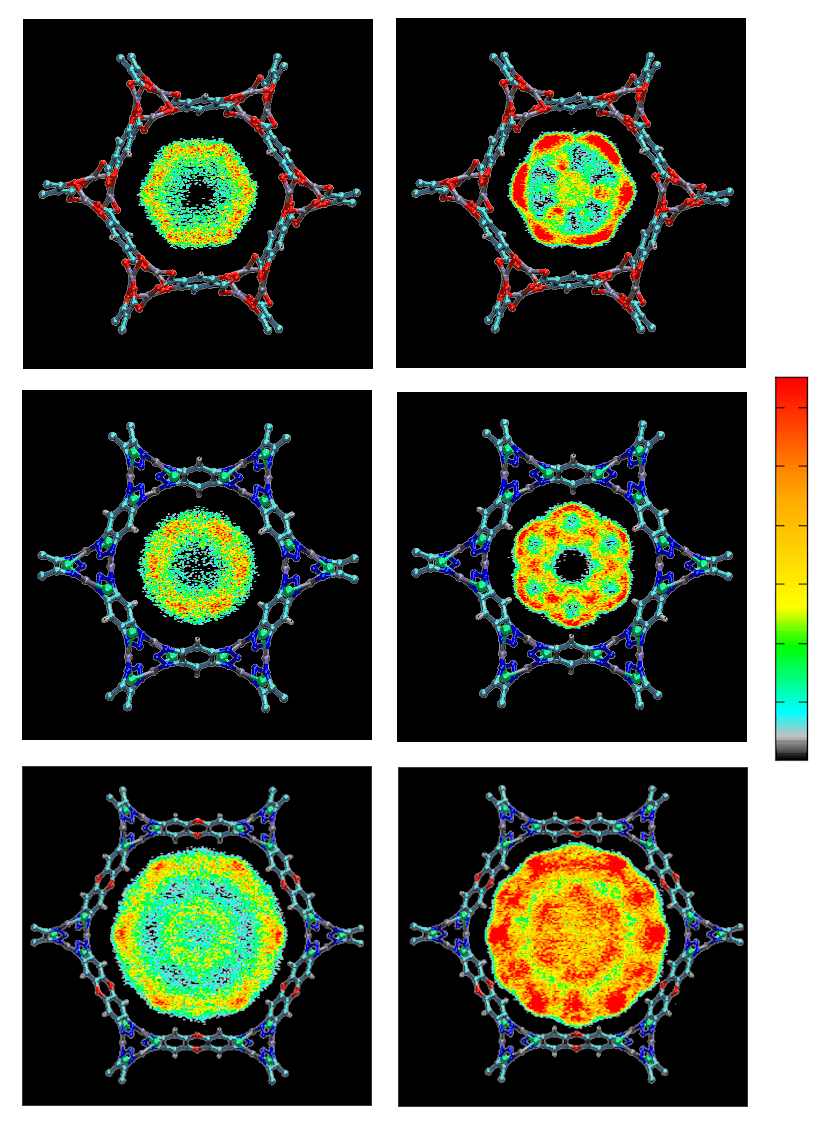}
\caption{Average occupation profiles of cyclohexane (left) and benzene (right) in Ni-MOF-74 (top), Ni-ClBBTA (center), and Ni-ClBTDD (bottom) at 298K and saturation conditions.}
\label{label:fig_07}
\end{center}
\end{figure}

Another critical factor in achieving efficient separation is the pore-filling mechanism. The selected MOFs have a high density of open metal sites accessible to molecules through the hexagonal channels. The steep adsorption isotherms of benzene suggest that, after reaching the onset pressure values, the adsorbed molecules occupy all the metal centers. With increasing pressure, benzene molecules fill the remaining accessible volume in the center of the pore. However, cyclohexane molecules do not entirely fill the cavities at pressures lower than their vapor saturation pressure. Figure \ref{label:fig_06} illustrates this pore-filling mechanism for Ni-MOF-74. The aforementioned pore-filling differences between benzene and cyclohexane can be extrapolated to the other two MOFs. Figure \ref{label:fig_07} displays the average occupation profiles of the adsorbates in the MOFs under saturation conditions. In all cases, adsorbed benzene molecules are distributed across additional adsorption sites, resulting in a higher density than cyclohexane when confined within the pores. When comparing the three MOFs, similar patterns are observed in the density profiles of cyclohexane. However, these patterns are slightly different for benzene, indicating that the conformation of the molecules adsorbed close to the internal walls influences the distribution of molecules in the center of the cavities.

\begin{figure}[!t]
\begin{center}
\includegraphics[width=1.0\linewidth]{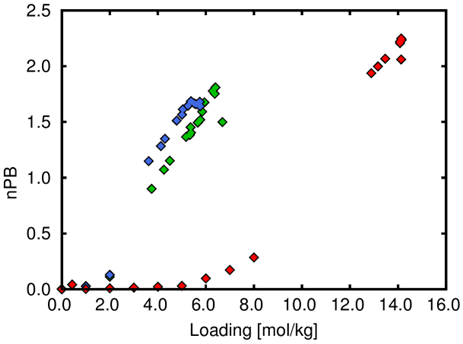}
\caption{Average number of $\pi$ bonds per molecule of benzene as a function of the loading in Ni-MOF-74 (green), Ni-ClBBTA (blue), and Ni-ClBTDD (red) at 298K.}
\label{label:fig_08}
\end{center}
\end{figure}

It is known that guest-guest intermolecular interactions for benzene are governed by $\pi-\pi$ interactions through the aromatic rings. This type of interaction is characteristic of aromatic compounds due to the presence of the electronic cloud of the ring. As observed in Figure \ref{label:fig_08}, a higher number of benzene molecules inside the framework implies a stronger interaction between them. This interaction increases more rapidly for adsorbents with smaller pore sizes. Ni-ClBTDD shows the highest interaction value because it is the framework with the highest loading capacity, making the aggregation of benzene molecules favorable. This fact allows benzene molecules to be adsorbed in the center of the pore, completely filling the available space, as shown in Figure \ref{label:fig_07}. A similar representation cannot be provided for cyclohexane molecules due to the absence of an aromatic ring.

In summary, during the competitive co-adsorption of benzene and cyclohexane, benzene molecules tend to occupy all the accessible metal centers of the adsorbents before cyclohexane can be adsorbed in the structure. Once all the metal centers are filled, benzene molecules can aggregate through $\pi-\pi$ interactions, populating the center of the cavities too. This hinders the adsorption of cyclohexane, which is excluded from the mixture. These results highlight the potential of large cavity MOFs with open metal sites to effectively separate these two aromatic molecules.

\section{Conclusions}
\label{sec:conclusions}

Computational techniques were employed to investigate the adsorption-based separation of benzene and cyclohexane using metal-organic frameworks with remarkable loading capacities. Monte Carlo simulations were utilized to study the pure and competitive adsorption of these two volatile organic compounds in metal-organic frameworks. Three adsorbents with similar topology, featuring open metal sites and pore sizes ranging between 10 and 20 Å, namely Ni-MOF-74, Ni-ClBBTA, and Ni-ClBTDD, were selected. Comparative analysis with literature data revealed that MOFs with large cavities and open metal sites exhibit high separation performance of benzene over cyclohexane.

To elucidate the molecular mechanisms involved in this separation, we first studied the low-coverage interactions of the guest molecules with the adsorbents, followed by the pore-filling mechanisms. As anticipated, the presence of open metal sites is crucial for the adsorption of the initial benzene molecules. However, this also has implications for the separation performance at saturation conditions. Once the benzene molecules cover the internal surface of the hexagonal pores of these MOFs, they can nucleate in the center of the cavity through $\pi-\pi$ interactions. Consequently, benzene molecules occupy the accessible pore of the MOFs, excluding cyclohexane from the initial mixture and resulting in high adsorption selectivity values. Our results underscore the potential of the selected MOFs to effectively separate benzene and cyclohexane, surpassing the separation capabilities of diverse porous materials. Notably, Ni-ClBTDD exhibits very high separation performance and adsorption capacity, both essential for designing efficient separation processes.


\section*{Supporting Information}

Supporting Information available.

\section*{Acknowledgements}

This work was supported by the Spanish “Ministerio de Educación, Cultura y Deporte” with a predoctoral fellowship (FPU16/04322). We thank C3UPO for the HPC support. 

\section*{Author contributions}

Here the author contributions.

\section*{Competing interests}

The authors declare that they have no known competing financial interests or personal relationships that could have appeared to influence the work reported in this paper.

\bibliography{Arxiv-Bz-cC6-OMS-MOFs}





\newpage

\renewcommand{\figurename}{Figure.}
\renewcommand{\thetable}{S\arabic{table}}  
\renewcommand{\thefigure}{S\arabic{figure}} 
\setcounter{figure}{0}

\onecolumngrid 

\newpage

\begin{center}
    \textbf{\Huge{Supporting Information}}

    \vspace{1.0cm}

    for

    \vspace{1.0cm}

    \textbf{\Large{Understanding the Role of Open Metal Sites in MOFs for the Efficient Separation of Benzene/Cyclohexane Mixtures}}
\end{center}

\begin{figure}[h!]
    \centering
    \includegraphics[width=0.75\textwidth]{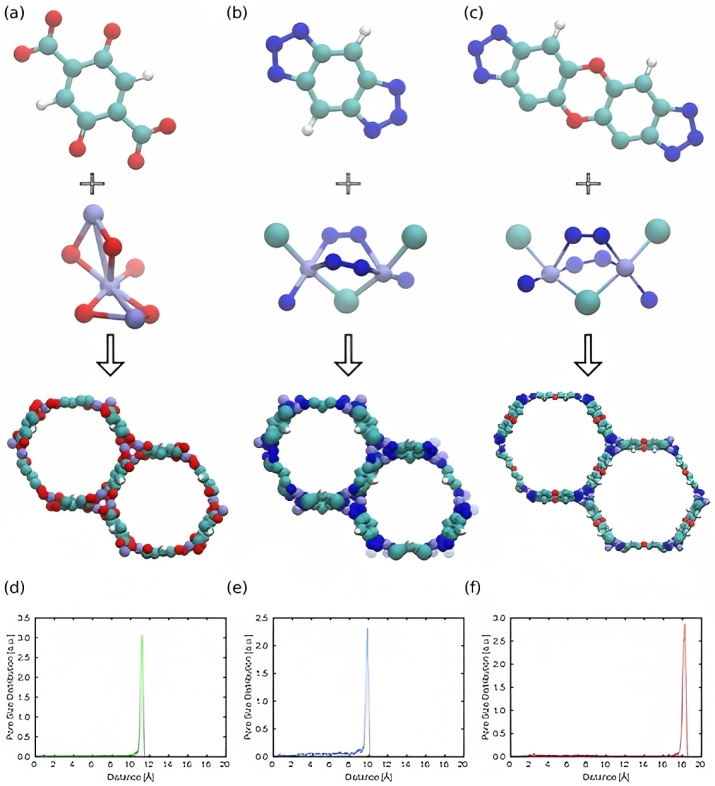}
    \caption{ Organic ligand, metallic cluster, and structure of (a) Ni-MOF-74, (b) Ni-ClBBTA and (c) Ni-ClBTDD. Pore size distribution of (d) Ni-MOF-74, (e) Ni-ClBBTA and (f) Ni-ClBTDD.}
    \label{fig:fig_S1}
\end{figure}

\newpage

\begin{figure}[h!]
    \centering
    \includegraphics[width=0.85\textwidth]{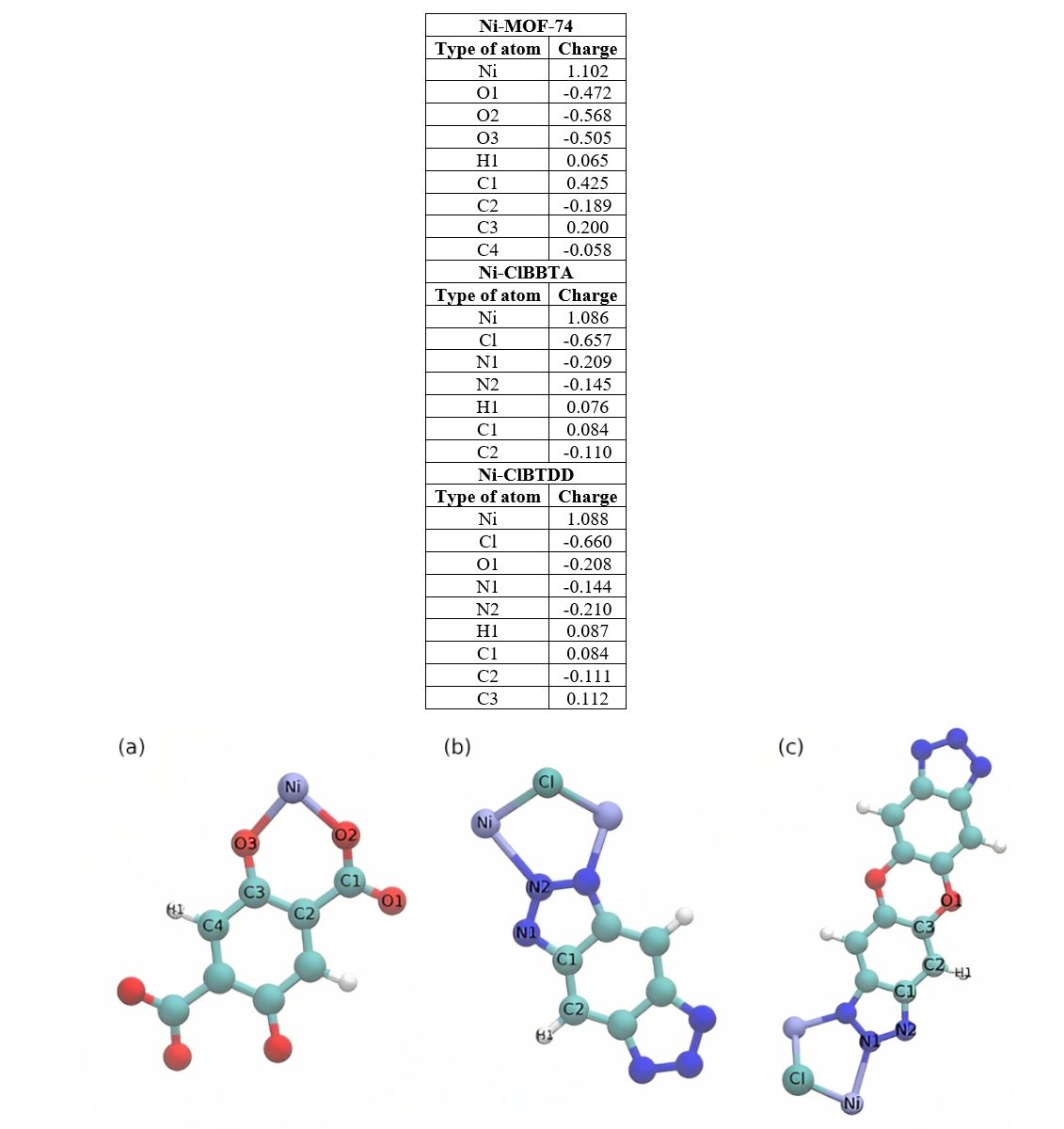}
    \caption{ Atomic charges of (a) Ni-MOF-74, (b) Ni-ClBBTA and (c) Ni-ClBTDD.}
    \label{fig:fig_S2}
\end{figure}

\newpage

\noindent Table S1. Dual-site Langmuir Sips fitting parameters for pure component adsorption of benzene and cyclohexane in the selected adsorbents.

\begin{figure}[h!]
    \centering
    \includegraphics[width=0.6\textwidth]{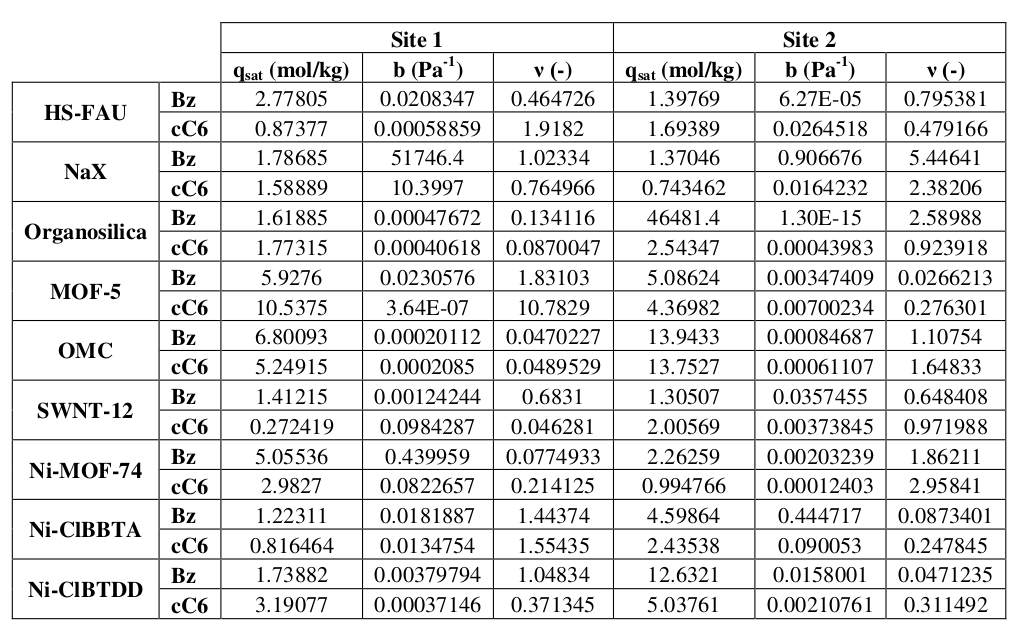}
\end{figure}

\noindent Table S2. P(Bz/cC6), Productivity of the separation and $\alpha$ Bz, separation factor as the trade-off between selectivity and adsorption capacity for the equimolar mixture of benzene and cyclohexane in the selected adsorbents at 10 kPa and 298 K. 

\begin{figure}[h!]
    \centering
    \includegraphics[width=0.4\textwidth]{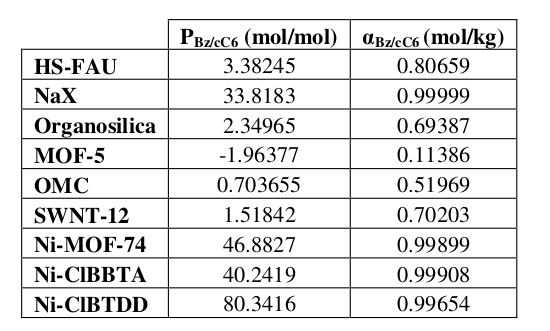}
\end{figure}
   
\noindent Table S3. P(Bz/cC6), Productivity of the separation and $\alpha$ Bz, separation factor as the trade-off between selectivity and adsorption capacity for the equimolar mixture of benzene and cyclohexane calculated from the adsorption equilibrium reported by Jansen et al.[1] at 10 kPa and 293 K. 

\begin{figure}[h!]
    \centering
    \includegraphics[width=0.5\textwidth]{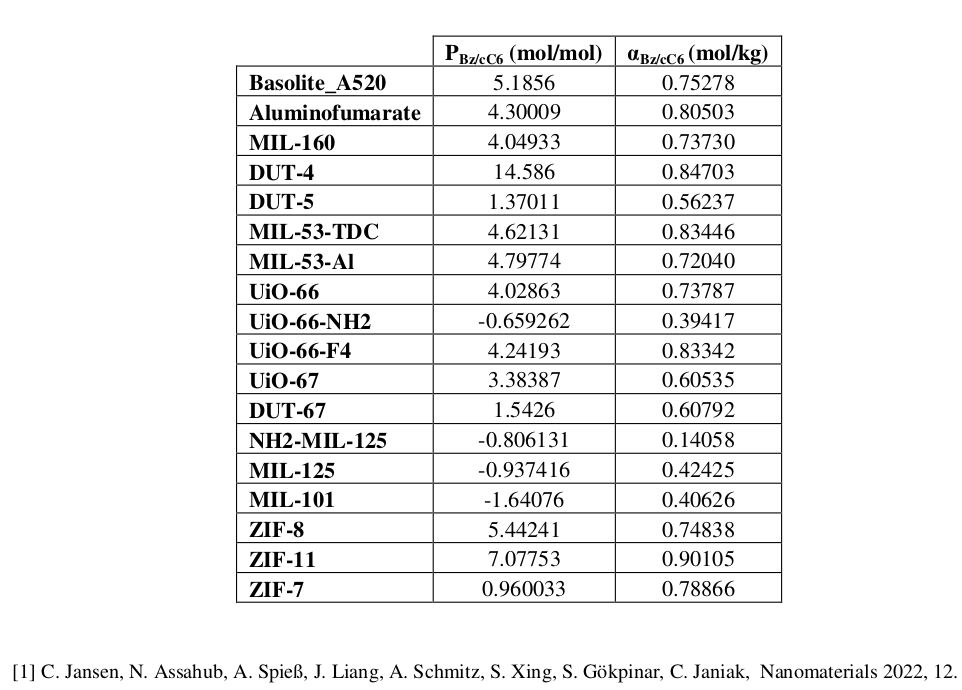}
\end{figure}

\end{document}